# Electron impact ionization of atoms

B. Tsipinyuk, A. Bekkerman and E. Kolodney

Department of Chemistry, Technion – Israel Institute of Technology

32000 Haifa, Israel

**Abstract**

A one parameter expression for the single ionization cross-section of atoms by electron impact is presented. Using this expression, the agreement obtained with available experimental data for 45 elements (from ionization threshold up to 200 eV ionizing electron energy) is shown to be comparable with (or better than) that achieved using other multi-parameter approaches. We suggest that the single parameter used here is associated with the effective reduction in the number of equivalent electrons within a given shell accessible for ionization by electron impact. We attribute this reduction effect to intra-shell shadowing of part of the shell electrons by other electrons of the same shell. For atoms with $2s^2p^i$ ($i$ = 3,4,5,6) outer shell configurations we suggest also an intershell shadowing effect. Finally, we discuss the possibility of a meaningful contribution of inner shell ionization to the single ionization cross-section due to various post-collision interactions. These interactions may be responsible for decreasing the formation probability of doubly charged ions.

## 1. Introduction

Electron impact ionization of atoms and molecules was studied during the last hundred years, from the very early days of modern physics (see Reviews [1,2]) .Nevertheless, the interest in this field has hot weakened also during the last decade. The recent theoretical studies of Kim and Rudde and co-workers [3-7] and Deutsch and Märk and co-workers [8-12], as well as of Rost [13], Saksena et al. [14], Khare et al. [15] and Joshipura et al. [15-16] should be mentioned along with numerous experimental studies (see, for example Refs. [18-27]).An important driving force for these studies are various applications such as electron impact ionization mass-spectrometry, simulation of plasma related phenomena, astrophysics and astrochemistry etc.

The major challenge in the field of electron impact ionization of atoms is to develop a general theoretical framework, which will provide an accurate ionization cross-section for many atoms over a practically relevant impact energy range. Due to its complexity, the fully quantum-mechanical treatment of electron impact ionization of atoms is possible only for the simplest cases of hydrogen and helium. Even in these cases the accuracy in the low energy range of $E \leq 100\ E_{n\ell}$ is quite limited (with E as the kinetic energy of the ionizing (primary) electron and $E_{n\ell}$ as the bond energy (absolute value of the full energy) of the atomic electron in the ($n\ell$) shell) [1]. With regards to heavier atoms, very recently Joshipura et al [16,17] have reported a quantum-mechanical numerical calculation of a *total* ionization cross-section $\sigma_{ion}$ for electron impact ionization of halogen atoms. However, the calculations were actually of the total inelastic cross-section $\sigma_{inel}$ (including ionization and excitation for all allowed electronic channels). The transition from $\sigma_{inel}(E)$ to $\sigma_{ion}(E)$ was done by multiplying $\sigma_{inel}(E)$ by some function $R(E)$. The specific form of this function was chosen to provide best agreement of $\sigma_{ion}(E)$ with experiment [17].

Given the limited success of the fully quantum treatment it turns out that various classical and semi-classical approaches are more fruitful. In the following background section we review some semiclassical approaches aimed at calculating *single* ionization cross-section $\sigma^+$. The degree of agreement between calculated and experimental $\sigma^+(E)$ dependences is presented and the predictive power of the different approaches is discussed. We try to assess advantages and disadvantages of the different approaches. In section 3 we present the central part of the paper, a single fitting parameter expression, which gives a satisfactory agreement between calculated and experimental $\sigma^+(E)$ dependences for 45 elements. Finally, in section 4 we consider the possibility of a meaningful contribution of inner shell ionization to the single ionization cross-section in terms of the (so-called) post-collision interactions.

## 2. Background

Let us assume that within the classical approximation the single ionization cross-section $\sigma_{nl}^+$ of the atomic electron shell ($nl$) is defined by only two parameters: the bond energy $E_{nl}$ and the ratio $E/E_{nl}$. From dimensional analysis (in atomic units) one then gets a general expression for $\sigma_{nl}^+$ in the form:

$$\sigma_{nl}^+ = \left(E_{nl}^{-2}\right) f\left(E/E_{nl}\right), \qquad (1)$$

where $f(E/E_{nl})$ is some function of the ratio $E/E_{nl}$. Converting this expression to ordinary units we obtain:

$$\sigma_{nl}^+ = 4\pi a_0^2 \left(\frac{Ry}{E_{nl}}\right)^2 f\left(E/E_{nl}\right), \qquad (2)$$

with Ry as the ionization potential (IP) of a hydrogen atom and $a_o$ as the Bohr radius. One should note that the factor π, which was included due to azimuthal symmetry of the cross-section, could be introduced inside the function $f$. The overall cross-section $\sigma^+$ for single ionization of an atom will be:

$$\sigma^+(E) = \sum_{n,l} N_{nl}\, \sigma_{nl}^+(E), \qquad (3)$$

where $N_{nl}$ is the number of equivalent electrons in the ($n, l$) electron shell. Using Eq. (2) we can write

$$\sigma^+(E) = 4\pi a_0^2 \sum_{n,l} N_{n,l} \left(\frac{Ry}{E_{nl}}\right)^2 f\left(E/E_{nl}\right). \qquad (4)$$

The $\sigma^+(E)$ dependence was calculated within the binary encounter model for the electron-atom collision, using both classical and semiclassical approaches [28-34]. However, one should note that in Ref. [29-34] the atomic electron was characterized not only by the absolute value of its total energy $E_{nl}$, but also by an additional parameter, its average kinetic energy $E_{nl}^k$. Namely, the function $f$ in Eqs. (2) and (4) is actually a function of two energy ratios: $f \equiv f\left(E/E_{nl}, E_{nl}^k/E_{nl}\right)$. Only for the case of the hydrogen atom the equality $E_{nl}^k = E_{nl}$ holds. However, Gryzinski [29] used this equality for all atoms such that in the final expression for $\sigma^+(E)$ the function $f$ depends only on the ratio $E/E_{nl}$ as in Eqs. (2) and (4). Burgess [32, 33] and Vriens [35-37] combined the classical binary encounter approximation with a quantum effect namely an exchange interaction (similar to Mott's treatment of this effect [38, 39]). They have argued [32, 33, 35-37] that the primary (incoming) electron with initial energy $E$ gains a kinetic energy $W$ <u>before</u> it interacts with the atomic electron. The function $f\left(E/E_{nl}, E_{nl}^k/E_{nl}\right)$ derived by these authors contains the general factor $E_{nl}/E$ in which $E$ is replaced by $E+W$. Burgess [32, 33] used $W = E_{nl}$ while Vriens [35-37] argued that $W = E_{nl}^k + E_{nl}$. Since then, this so-called "acceleration correction" was adopted by others as well and introduced into several electron impact ionization models (see, for example [4-6, 34]). We think that this transition from $E$ to $E+(E_{nl}^k + E_{nl})$ overestimates the "acceleration" effect. The question rises due to the vague meaning of the expression "<u>before</u> primary electron interacts (or collides) with the atomic electron", specifically, the ill-defined nature of the word "before" in this context.

    To clarify this issue, let us consider the simplest case of hydrogen atom ionization. Vriens [35-37] and Ochkur [34] assumed that collision occurs at the crossing of the primary electron trajectory and the orbit of the atomic electron. For the ground state hydrogen atom they used $W = E_{1s}^k + E_{1s} = 2Ry$, namely here $W$ is the absolute value of the potential energy of the atomic electron in the field of the nucleus. For the sake of clarity, let us define the phrase "before collision" in terms of the relative position of the incoming electron with respect to the atomic electron, such that absolute value of their interaction potential energy $|U|$ is much smaller than the bond energy $E_{nl}$ of the atomic electron. For the hydrogen atom it means $|U|<<Ry$. The condition $|U|\sim Ry$ holds when the incoming electron crosses a spherical surface with radius $2a_o$ ($0.67 Ry < |U| < 2Ry$). One can therefore say that electron – electron collision occurs already at the $2a_o$ radius. Consequently, the condition $|U|<< Ry$ holds when the distance between the incoming electron and the nucleus is much larger than $2a_o$. However, at this distance the energy gain of the primary electron "before collision" with the atomic electron will be much smaller than $e^2/2a_o \equiv Ry$ and certainly much smaller than $(E_{1s} + E_{1s}^k) = 2Ry$. For the general case it means that the energy gain "before collision" with the outer shell electron has to

be much smaller than $E_{nl} + E_{nl}^k$. Indeed Ochkur [34] and Kim *et al* [4-6] (see below) empirically found that in order to improve the agreement between theory and experiment (for multi-electron atoms) one has to decrease the value of $E_{nl}^k$ by a factor of 4-5 [Ref. 34] or by a factor of $n$ [Ref. 4-6], that is practically to set $E_{nl}^k \approx E_{nl}$ as was done by Gryzinski [29]. We think that the argumentation given above is providing the physical basis for their empirical conclusion.

We will present here the expression for $f(E/E_{nl})$ which was derived by Gryzinski [29]. As will be shown later, this expression leads in some cases to a rather good agreement between calculated and measured $\sigma^+(E)$ dependences. This expression is given by:

$$f_G\left(x \equiv E/E_{nl}\right) = \frac{1}{x}\left[\frac{x-1}{x+1}\right]^{\frac{3}{2}}\left[1 + \frac{2}{3}\left(1 - \frac{1}{2x}\right)\ln\left(2.7 + \sqrt{x-1}\right)\right]. \quad (5)$$

From this point and on the function $f_G(x)$ will be called the Gryzinski function, while Eq. (4) using $f_G(x)$ - the Gryzinski formula.

In the limit of high electron energies $E$ the generally accepted theory of atomic ionization is the quantum asymptotical ($E/E_{nl} \gg 1$) theory by Bethe [40] (see also ref. [1]) that results in the following expression for $\sigma_{nl}^+$:

$$\sigma_{nl}^+(E) = \pi \left\langle r_{nl}^2 \right\rangle f_B\left(E, E_n\right), \quad (6)$$

with the Bethe function

$$f_B\left(E, E_{nl}\right) = (C/E)\ln\left(4\tau E/E_{nl}\right), \quad (7)$$

where $C = (8/3)Ry$, $\tau$ has a numerical value smaller than one and $\left\langle r_{n\ell}^2 \right\rangle$ is the mean square radius of the ($nl$) shell. From Eq. (7) one can see that the Bethe function $f_B(E, E_{nl})$ is not a function of $E/E_{nl}$ alone but also includes the factor $E^{-1}$ common for all shells. Mann [41] suggested to extend the use of the Bethe approximation to the range of small E values down to $E = E_{nl}$. He has simplified and modified expressions (6) and (7) by putting $4\tau = 1$ and introducing the fitting parameters $A_l$ and $p$, such that

$$\sigma_{nl}^+(E) = A_l \left\langle r_{nl}^2 \right\rangle (1/E)^p \ln\left(E/E_{nl}\right)^p, \quad (8)$$

where $p=1$ for $l=0$ and $p=3/4$ for $l \geq 1$. Please note that the fitting parameter $A_l$ in Eq. (8) has dimension of [energy $^p$] and therefore depends on $l$. However, it seems that Mann choose to use this factor with the same numerical value $A$ for all $l$, such that

$$\sigma^+(E) = A\sum_{n,l} N_{nl} \left\langle r_{nl}^2 \right\rangle \left(\frac{1}{E_{nl}}\right)^p \frac{\ln x^p}{x^p}, \quad (9)$$

with $x \equiv E/E_{nl}$. The $A$ value was derived by equating the maximal value of $\sigma^+$ calculated by Eq. (9) to the corresponding measured value for electron impact single ionization of argon ($2.83 \times 10^{-16}$ cm² [41]). If one measure $\langle r_{nl}^2 \rangle$ in Å² and $E$ in eV, then $A=10.32 \times 10^{-16}$ eV for $l = 0$ and has the same numerical value (but in units of eV$^{3/4}$) for $l \geq 1$.

Kim et al. [3-6] developed the binary-encounter-dipole (BED) model for electron impact ionization of atoms and molecules. They combined the modified Vrien's symmetric binary encounter theory [35-37] including interference between the direct and the exchange terms and the Bethe theory [40] for fast incident electrons. Their final formula for $\sigma_{nl}^+(E)$ does not contain any fitting parameter but accurate experimental or theoretical data on photoionization cross-sections of $(n, l)$-shells are required in order to derive the differential dipole oscillator strength for ionization. The BED theory gives very good agreement with experimental $\sigma^+(E)$ dependences for several light atoms (H, He, Ne) in the range of IP $\leq E \leq 10^4$ eV but due to the lack of accurate data on the differential oscillator strength for the majority of multi-electron atoms it is practically applicable only to a very limited number of atoms. For the case that nothing is known about the differential or total dipole oscillator strengths a simplified model (the Binary-encounter-Bethe (BEB) model) was proposed [3]. This BEB model, which was applied also for some multi-electron atoms [6,7], "leads to ionization cross sections of correct orders of magnitude" [3].

We would like to make one comment regarding the BED and BEB models:

Kim and Rudd have elegantly shown [3] that the differential ionization cross-section $d\sigma/dw$ (energy distribution of electrons after an electron – atom ionizing collision) can be represented as the sum of three pairs of some functions $\sum_{i=1}^{3} F_i(t)[f_i(w+1) + f_i(t-w)]$ with relative energies $t = E/E_{nl}$ and $w = E_2/E_{nl}$ ($E_2$ is the ejected electron energy). The functions $f_i(w+1)$ representing the ejected electrons, $f_i(t-w)$ representing the scattered electrons and their sums are symmetric in $w$ within the $[0, t-1]$ interval relative to its center. This symmetry is due to the indistinguishability of ejected and scattered electrons after collision. The integrated (over $w$) cross section can be obtained either by integration of each pair of the sum over the $[0, (t-1)/2]$ interval or by integration of only one term (irrespective whether the first or the second one) of each (or any) pair in the sum within the $[0, t-1]$ interval (leaving out the other term of the pair). For example, the cross section $\sigma^+(E)$ in the binary-encounter-Bethe (BEB) model (formula (57) in ref. [3]) could be obtained either by the integrals $\sum F_i(t) \int_0^{(t-1)/2} [f_i(w+1) + f_i(t-w)]dw$, or by the integrals $\sum F_i(t) \int_0^{t-1} f_i(w+1)dw$ only, or by the integrals $\sum F_i(t) \int_0^{t-1} f_i(t-w)dw$ only. Regarding the binary-encounter-dipole (BED) model (formula (55) in Ref. [3]), since the $F_3(t)f_3(t-w)$ term in this model was left out [3] the integration of the other symmetric term $F_3(t)f_3(w+1)$ of the third pair in the sum must be carried out over the interval $[0, t-1]$. Consequently, in Eq. (56) of Ref.[3] for $D(t) \equiv \int f_3(w+1)dw$ the upper limit of the integral must be $t-1$ and not $(t-1)/2$ (see also [15, 42]). This change results in some underestimation of calculated BED cross-sections $\sigma^+(E)$ relative to these presented in [3] for H, He, Ne and H$_2$ in the range of low $E$ energies.

Deutsch and Märk [43] introduced a modification of formula (4). It was suggested, quoting from ref. [43], "to replace $4a_0^2$ by $g_l(r_{max})_{nl}^2$, and in addition, to remove the term $(Ry/E_{nl})^2$", where $(r_{max})_{nl}$ is the radius of maximum charge density of the (nl) shell and $g_l$ is a fitting factor (named a weighing factor in ref. [43]) depending only on orbital quantum number $l$. The modified Eq. (4) (with $f_G$) now takes the form:

$$\sigma^+(E) = \pi \sum N_{nl}\, g_l\, (r_{max})_{nl}^2\, f_G(E/E_{nl}). \qquad (10)$$

It was suggested to set $g_s = 3$ (s electrons) and $g_l = 0.5$ (electrons with $l \neq 0$) independent of the electron bond energy $E_{nl}$. The contribution of s-electrons to $\sigma^+$ is therefore 6-fold as compared to $l \neq 0$ electrons (for a given $r_{max}$ value). As will be shown later, the $\sigma^+(E)$ dependencies calculated according to Eq. (10) for a number of elements agree with the measured ones rather poorly. To improve agreement with experiment Margreith *et al.* [8] suggested to treat the weighing factor $g$ in Eq. (10) as depending not only on $l$ but also on $n$, $N_{nl}$ and $E_{nl}$ such that the $g$ factor used is given by

$$g(n, l, N_{nl}, E_{nl}) = \frac{C(n, l, N_{nl})}{E_{nl}}, \qquad (11)$$

where the value of $C(n,l,N_{nl})$ is considered to be a free fitting parameter. In addition, the Gryzinski function $f_G(x)$ (Eq. (5)) was modified by using additional fitting parameters (see also ref. [11]) which change its shape and maximum value, such that

$$f_G(x, a, b, c, d) = \frac{d}{x}\left[\frac{x-1}{x+1}\right]^a \left[b + c\left(1 - \frac{1}{2x}\right)\ln\left(2.7 + \sqrt{x-1}\right)\right]. \qquad (12)$$

For each of the *s*, *p*, *d* and *f* orbitals a different set of *a*, *b*, *c*, *d* parameters was fitted. Eq. (10) could therefore be rewritten as

$$\sigma^+(E) = \pi \sum_{n,l} N_{nl} (r_{max})_{nl}^2 \cdot \frac{C(n,l,N_{nl})}{E_{nl}} \cdot f_G(E/E_{nl}, a, b, c, d), \qquad (13)$$

where the *a*, *b*, *c*, *d* parameters depend on *l*. For the calculation of $\sigma^+(E)$ dependencies for 45 elements (refs. [8, 11]), a matrix with 50 fitting parameters $C(n,l,N_{nl})$ was used. Combined with 4 sets of *a*, *b*, *c*, *d* parameters for *s*, *p*, *d* and *f* electrons the number of fitting parameters exceeded 60. The authors of Refs. [8,11,43] named this multi-parameter approach the DM-formalism[*].

In what follows we will present a single parameter expression, which will allow us to describe experimental $\sigma^+(E)$ curves for 45 elements. The agreement achieved between the calculated and measured $\sigma^+(E)$ curves using our method is comparable to (not worse than) that achieved by the multi-parameter DM-formalism.

## 3. Intrashell and intershell shadowing. One-parameter expression for the calculation of σ⁺(E).

In the asymptotical formula of Bethe (6) the mean square radius $\langle r_{nl}^2 \rangle$ is a factor that defines the value of the partial ionization cross-section $\sigma_{nl}^+$ for given values of E and the $E/E_{nl}$ ratio. Going to the range of small E values one can expect that although the $\sigma_{nl}^+(E)$ dependence will be different than the Bethe $E^{-1}f_B(E/E_{nl})$ dependence (see Eq. (7)), the value of $\sigma_{nl}^+$ will still be defined by the factor $\langle r_{nl}^2 \rangle$ rather than by the factor $(r_{max})_{nl}^2$ as, for example, in the DM-formalism (Eqs. (10), (13)). Moreover, we will assume that the factors $\langle r_{nl}^2 \rangle$ and $N_{nl}$ are the basic factors that jointly with the function $f(E/E_{nl})$ define the partial cross-sections $\sigma_{nl}^+(E)$ and consequently the total cross-section $\sigma^+(E)$:

$$\sigma^+(E) = \pi \sum_{nl} N_{nl}^{eff} \langle r_{nl}^2 \rangle f_G(E/E_{nl}), \qquad (14)$$

with the Gryzinski function $f_G(E/E_{nl})$ used for the function $f$. The factor $N_{nl}^{eff}$ in formula (14) is the effective number of equivalent electrons in the (nl)-shell accessible for ionization by impact of primary electron. We assume that the effective number $N_{nl}^{eff}$ is smaller than the number $N_{nl}$ due to shadowing of part of the shell electrons by other electrons of the same shell (intrashell shadowing) or by electrons of another (outer) shell (intershell shadowing).

---

[*] Only in the first variant of the DM formalism [43], the authors correctly define the $(r_{max})_{nl}$ in formula (10) ($r_n$ in [43]) as radius of maximum charge density in line with the numerical values that were used in ref. [43]. However, in the following numerous publications (see for example [9, 11, 12]) using as before the values of radius of maximum charge density from Desclaux tables [44], the authors incorrectly named these values as root mean square radii (or $r_{nl}^2$ as mean square radius). Please note that if one is using in formula (13) the mean square radius $\langle r^2 \rangle_{nl}$ instead of the square of radius of maximum charge density $(r_{max})_{nl}^2$ (both from the Tables in [44]) together with the values of fitting parameters $C(N, l, N_{nl})$ from table in [8] or table 1 in [11] or [12], then the calculated values of $\sigma^+(E)$ will be overestimated compared with the experimental ones by a factors of 1.5 ÷ 3.

Let us first consider the intrashell shadowing effect. From a classical point of view, the "movement" of the equivalent electrons in the shell is being constantly correlated, such that at each instant half of the electrons in a filled shell reside on the "frontal hemisphere" (as "seen" by the approaching primary electron) and thus are screening (for the primary electron) the second half of the electrons in the "back hemisphere". In the limit of *full* shadowing of the back electrons, only frontal electrons (e.g., half of the filled shell electrons) are subjected to ionization by the incoming electron, resulting in a singly-ionized atom. In this case $N_{nl}^{eff} = N_{nl}/2$. In all other cases when shadowing is not complete, one should get $N_{nl}/2 < N_{nl}^{eff} < N_{nl}$.

We would like to discuss now the intrashell shadowing for different shells. With regard to filled $ns^2$-shells, the correlated movement of the two electrons means that appreciable shadowing of one electron by the other can take place only for the unique orientation where both of them are situated along the approach line of the primary electron (or close to it). Since such orientations constitute a small part of all possible orientations, the effect of intra-shell shadowing for $ns^2$ shells can be neglected. Following similar arguments, we choose to neglect intra-shell shadowing effects for *nl*-shells with $N_{nl} \leq 4$. Namely, for $N_{nl} \leq 4$ we use $N_{nl}^{eff} = N_{nl}$ for all *l*. To estimate the intra-shell shadowing for *nl*-shells with $N_{nl} > 4$ we will use a rather crude model of shadowing. For a filled shell one can assume that ($N_{nl}/2$) electrons out of $N_{nl}$ always reside on the frontal hemisphere (as viewed by the incoming electron) and therefore are not shadowed. At the same time, each of these ($N_{nl}/2$) electrons shadows a certain part $\beta$ of each of the ($N_{nl}/2$) electrons in the back hemisphere. The effective number of electrons $N_{np}^{eff}$ accessible for ionization by the primary electron can then be given by the simple expression:

$$N_{np}^{eff} = N_{nl}\left(\frac{1}{2} + \frac{1}{2}(1-\beta)^{N_{nl}/2}\right); \qquad N_{nl} > 4. \qquad (15)$$

In this expression the first term is the number of electrons in the frontal hemisphere of (nl)-shell and the second one is the number of electrons in the back hemisphere accessible for ionization by impact of primary electron.

The issue of intershell shadowing is more complicated. One can assume that, if the corresponding values of $r_{max}$ (or $<r>$ or $\sqrt{<r^2>}$) for two shells are substantially different, the intershell shadowing of inner shell electrons by the outer shell electrons is considerably smaller than the intrashell (of the same inner shell) shadowing. Therefore, in the following we neglect intershell shadowing for the most contributing (outer) shells of all atoms except for the atoms with electron configurations $2s^2p^i$ (i = 3 – 6): N, O, F and Ne. For these elements, according to Desclaux's tables [44], the values of $r_{max}$ of 2s electrons is larger by 3-8% than $r_{max}$ of 2p electrons, while the values of $<r>$ (or $\sqrt{<r^2>}$) of 2s electrons is smaller by 8% (or 13%) than the corresponding values of 2p electrons. So one can consider (within the framework of the shadowing concept) (s + p)- electrons as belonging to the same shell and thus derive the $N_{2s}^{eff}$ and $N_{2p}^{eff}$ numbers using Eq. (15) for intrashell shadowing. As the two s-electrons does not shadow each other, we will only take into account the shadowing of the one of the 2s-electrons by 2p-electrons if $N_{2p} > 4$ (as for intrashell shadowing) namely only for the F and Ne atoms. To find $N_{2s}^{eff}$ for these atoms, one should replace the power ($N_{2s}$/2) in the second term of Eq. (15) by ($N_{2p}$/2). To find $N_{2p}^{eff}$ (for atoms with $N_{2s} + N_{2p} > 4$) one should replace the power ($N_{2p}$/2) in the second term of Eq. (15) by [($N_{2p}$/2)+1] as half of the 2p electrons are being shadowed by the other half of the 2p electrons and additionally by one of the 2s electrons.

The value of $N_{nl}^{eff}$ therefore equals $N_{nl}$ except for filled or nearly filled shells with $l \geq 1$, for example for $np^5$ and $np^6$ shells or $nd^6 \div nd^{10}$ shells. In these cases the effective number $N_{nl}^{eff}$ is defined by Eq. (15) in which the value of $\beta$ is treated as the only fitting parameter. As discussed in the next section the choice of $\beta = 0.48$ in Eq. (15) resulted in a sufficiently good agreement between calculated and measured $\sigma^+(E)$ dependencies for most of the data published until now. Note, that for different nl- shells with $5 \leq N_{nl} \leq 14$ using this $\beta$ value in Eq. (15) is equivalent to the corresponding variation of the ratio $N_{nl}^{eff}/N_{nl}$ within a rather narrow range of 0.51 – 0.60.

## 4. Comparison between measured and calculated $\sigma^+(E)$ dependences using different approaches.

In Fig. 1a, b, c, d, e we present the calculated $\sigma^+(E)$ curves ($E \leq 200$ eV) using Eq. (14) for 45 elements together with experimental results. Also presented, for the sake of comparison, calculated $\sigma^+(E)$ curves according to Gryzinski's formulae (4) and (5), Mann's equation (9) with three fitting parameters, and the two variants of the DM-formalism: Eq. (10) with two fitting parameters and the multi-parameter Eq. (13). The values of $E_{nl}$ for the outermost shells were taken from Moor's tables [45] and for the shells, next to the outermost ones, were also taken from [45] except the shells $np^6$ ($n$ =2,3,4.5 ) for Mg, Ca, Sr and Ba correspondingly and the shell $3d^8$ for Ni. For these shells and for the next inner shells of all elements the values of $E_{nl}$ were taken from Ref. [46]. For Yb and U the binding energies of all shells were taken also from Ref. [46]. The binding energies of $(n-1)s^2$ –electrons for alkali atoms (with ground state configuration $(n-1)s^2p^6ns$) were estimated as described in the Appendix.

Let us consider now the agreement of the calculations using the different approaches with the corresponding measured results. To a first approximation, the degree of this agreement will be given by comparing calculated and measured maximum cross-section values $\sigma^+_{max}$. In case that there are two or more experimental $\sigma^+(E)$ curves for a given element (all of them are shown in Fig. 1) we will evaluate the agreement limits (percentage of deviation from measured value) using the data that gives best agreement with calculation. First, we will consider the calculation of $\sigma^+(E)$ by Gryzinski (Eq. (4) with $f_G$ as in Eq. (5)), which is completely parameter free. We are not concerned here with the validity of the various approximations used in deriving the function $f_G$ (for this matter, see for example comments in refs. [1, 34]). As can be seen from Fig. 1, for 15 elements (H, Li, Na, Mg, P, K, Ca, Ga, Se, Rb, Te, Cs, Ba, Pb, Bi) the value of $\sigma^+_{max}$, calculated by the Gryzinski equations differs by no more than ~10% from the measured values. As for the rest of the elements, 14 of them are within (12-30)% of the measured values (He, Al, Si, S, Ge, As, Br, Kr, Sr, In, Sn, Sb, I, Xe), 7 elements are within 31-100% (C, N, Cl, Ar, Ti, V, Hg) and for 9 elements (O, F, Ne, Fe, Ni, Cu, Ag, Yb, U) $\sigma_{max}$ is overestimated by a factor of 2.4 ÷ 4.7.

As for Mann's formula (9) with the three-fitting parameters, a major drawback is the underestimated values of $\sigma^+_{max}$ for H and He (2.8 and 3.7 times correspondingly smaller than measured $\sigma^+_{max}$). But at the same time $\sigma^+_{max}$ values for 24 elements (O, F, Na, Al, S, Cl, Ar, K,

Ca, Fe, Ni, Cu, Ga, Br, Kr, Rb, Ag, I, Cs, Ba, Yb, Hg, Pb, Bi) are within ±10% agreement with measured values. This is much better than for the Gryzinski formula. 18 elements (Li, C, N, Ne, Mg, Si, P, Ti, V, Ge, As, Se, Sr, Sn, Sb, Te, I, Xe) lie within the (12-33)% agreement range. Very large overestimation (similar to that for Gryzinski formula) is found for uranium for which the calculated $\sigma^+_{max}$ value is 2.5 times larger than the measured one.

Next we will consider the calculated results using the DM-formalism. First, we discuss the two fitting parameters' variant [43] (Eq. (10)). The two free parameters (weighing factors $g_s$=3 and $g(l \geq 1)$=0.5) were fitted to obtain the best agreement between calculated and measured $\sigma^+(E)$ dependencies for ionization of noble gas atoms. Only for 15 elements (H, He, N, F, Ne, Na, Al, Ar, V, Cu, Kr, Ag, In, Xe and Pb) the calculated $\sigma^+_{max}$ values agree with the measured values to within ±10%. This is substantially worse than for Mann's formula but it should be mentioned that the last one has one additional fitting parameter.

Another characteristic feature of the first DM-variant is that for all elements with outer np-shells the calculated $\sigma^+(E)$ curve in the range of small E values (the rise region) is less steep than the measured one. Also, at the energy $E = E_{ns}$ for the nearest ns-subshell there is a rather sharp bend in the calculated $\sigma^+(E)$ dependence which is not observed experimentally. This bend is particularly distinct for the elements O, F, Ne, Cl, Ar, Se, Br, Kr (namely for the configuration $ns^2p^i$ with $i$=3, 4, 5, 6). This artificial feature clearly demonstrates that the proportion of weight factors $g_l$ for s- and p-electrons, as used in ref. [43], strongly overestimate the weight of s-electrons in the range of $\sigma^+_{max}$. This overestimation of $g_s$ is clearly manifested also in the overestimated $\sigma^+(E)$ for alkali and particularly (1.5-2 times larger) alkaline-earth metals. On the other hand, the underestimation of the $g_p$ factor leads to underestimation of calculated $\sigma^+_{max}$ by 15-40% for most elements with unfilled outer p-subshell.

The second multi-parameter variant of the DM-formalism (Eq. (13)) naturally yields the best agreement with experiment, as compared with the first DM variant, Gryzinski's formula and Mann's formula (at least according to the criterion of agreement between calculated and measured $\sigma^+_{max}$ values).

It is interesting to note that the multi-parameter DM-formalism is the only approach that provides satisfactory agreement between calculation and experiment for $\sigma^+(E)$ of uranium (50% overestimation of calculated $\sigma^+_{max}$, as compared with the measured one, while all other approaches are overestimating the measured $\sigma^+_{max}$ by factors of 2.5÷4.8 (see Fig. 2)). A closer inspection reveals that out of 6 $C(n, l, N_{nl})$ parameters used within the DM-formalism for the 6 electronic shells ($5d^{10}f^36s^2p^6d^17s^2$), the bond energies of which lie below 200 eV, 4 parameters are specific only for uranium. These parameters are $C(7, s, 2)$, $C(6, d, 1)$, $C(6, p, 6)$ and $C(5, f, 3)$ (see Eq. (13)). All these 4 parameters were chosen specifically for the corresponding shells of uranium and were not used for any other element.

Finally, we would like to compare the measured $\sigma^+(E)$ dependencies with the results of the calculation according to Eqs. (14) and (15) containing a single parameter $\beta = 0.48$ for filled or nearly filled ($N_{nl} > 4$) electron shells with $l \geq 1$. The calculated $\sigma^+(E)$ curves, according to this approach, are given in Fig. 1 together with other calculations using the other approaches and experimental results. According to the formerly used criterion of degree of agreement (percentage difference) between calculated and measured $\sigma^+_{max}$ values, we find that 35 elements (H, C, N, O, F, Na, Mg, Al, P, S, Cl, Ar, Ca, Ti, Fe, Ni, Cu, Ga, As, Se, Br, Kr, Sr, Ag, In, Sb, Te, I, Xe, Cs, Ba, Yb, Hg, Pb, Bi) fall within the ~10-12% agreement range. For some elements out of this list (H, C, Na, Mg, Al, P, S, Ti, As, Se, Te) and also for Li, He and Si (~20% difference between calculated and measured values of $\sigma^+_{max}$) the calculation of the $\sigma^+(E)$ dependence for the E range around $E_{max}$ is free of any adjustable parameter. For another 7 elements (Ne, K, Si, V, Ge, Rb, Sn) the agreement between calculated and measured values of $\sigma^+_{max}$ lies within the (~13÷30)% range. The largest difference (factor of 2) between calculated and measured value is found for uranium. One should note that this is the case also for the Gryzinski and Mann equations and the first variant of the DM-formalism (see Fig. 1e).

## 5. On the inner shell contribution to the atomic single ionization cross-section

The contributions $\sigma^+_{nl}(E)$ by inner shells with bond energies $E_{nl} \geq 50$ eV to the direct ionization cross-section are presented separately in Fig. 1. As can be seen, for several atoms (such as Fe, Ni, As, Se, Br, Kr, Te, I, Xe) taking into account single ionization of these shells results in considerable improvement of the agreement between the measured $\sigma^+(E)$ dependence and calculated one, as given for example by the expression of Mann (Eq. 9) or our one-parameter Eqs. (14)-(15). The better agreement is manifested by the shape of $\sigma^+(E)$, mainly the high-energy tail behavior.

However, a singly-charged ion formed following inner shell ionization is highly over-excited with respect to the inner-shell vacancy and therefore should be relaxed via an Auger process resulting in the formation of a doubly-charged ion. The calculated lifetime of $M_{45}$ vacancies in $Se^+$, $Br^+$, $Kr^+$ and $N_{45}$ vacancies in $Te^+$, $I^+$, $Xe^+$ is about $6.5 \cdot 10^{-15}$ sec. (with a corresponding level width $\Gamma \sim 0.1$ eV) [68]. During this time the singly charged ion (having a vacancy in its inner shell) cannot escape the ionization zone and therefore should be registered as a doubly charged ion. This conclusion implies that inner shell ionization cannot contribute to the cross-section of a singly charged ion production and is therefore in disagreement with our conclusion (see above) that inner shell ionization contributes meaningfully to single ionization cross section $\sigma^+$. In order to reconcile this (seemingly) contradiction, one has to assume the existence of some process that reduces the formation probability of the doubly-charged ions due to Auger relaxation of ions with inner shell vacancy. The effect of this hypothetical process has to be manifested also by a decreased yield of Auger electrons, as compared with that expected based on the ionization cross-section of the inner-shells under consideration (other possible mechanisms, such as autoionization, will be briefly discussed later).

The last conclusion can be examined using the data in Refs. [21,69-71] where the $\sigma_{nl}^+(E)$ dependence for the production of $M_{45}$ vacancies in Kr atoms and $N_{45}$ vacancies in Xe atoms are presented over the energy range which is of interest here. The $\sigma_{nl}^+(E)$ measurements in all these studies are based on measurements of the corresponding yields of Auger electrons. From the data presented in Refs. [21, 69-71] it is clear that the ionization cross-sections for 3d-shell of Kr and 4d-shell of Xe are much smaller than those predicted by Mann''s expression (8) and our one-parameter expression (Eq. (14)). At this point we would like to remind the reader that the last two expressions provide a rather good agreement with the experimental $\sigma^+(E)$ for transition metal atoms with a large contribution of outer 3d and 4d shells (see Fig. 1). For example, for $M_{45}$ vacancies of Kr, the experimental value is $\sigma_{3d}^+$ (200 eV) $\simeq 1.75 \cdot 10^{-18}$ cm$^2$ [69,70] while the calculated value of $\sigma_{3d}^+$ (200 eV) is $3.2 \cdot 10^{-17}$ cm$^2$ according to Eq. (14) or $1.1 \cdot 10^{-17}$ cm$^2$ according to Mann's expression (8). For $N_{45}$ vacancies in Xe the measured cross section is $\sigma_{4d}^+$ (200 eV) = $1.3 \cdot 10^{-17}$ cm$^2$ [21, 71], while the calculated value is $\sigma_{4d}^+$ (200 eV) = $9.2 \cdot 10^{-17}$ cm$^2$ according to Eq. (14) and $3.9 \cdot 10^{-17}$ cm$^2$ according to Mann's expression (8). Following these comparisons, one can conclude that indeed the measured yields of Auger electrons resulting from the ionization of 3d- and 4d-shells are substantially reduced as compared with the yields predicted by the ionization cross-sections of these shells calculated according to formulae (8) or (14).

We suggest that the kind of processes that may be responsible for the decrease in probability of Auger transitions (leading to the ejection of Auger electrons and generation of doubly-charged ions) are associated with a variety of post-collision interactions (PCI) [72-74]. During the process of ionization of the bound electron there is some probability that either the inelastically scattered electron or the directly ionized (atomic) electron will be moving so slowly that at the instant of the Auger transition this slow electron will still be at the vicinity of the atom. The subsequently emitted fast Auger electron will overtake this slow electron, which will suddenly be subjected to the filed of a doubly charged ion. The slow electron can then be recaptured, resulting in the formation of an excited (Rydberg state) singly-charged core. The probability for this process is larger in the near-threshold (for inner-shell ionization) range of the ionizing electron energy $E$. The above mechanism described in the literature for the interaction between the Auger electron and the slowly outgoing electron is experimentally manifested by a shift of the Auger peak (in the energy spectrum of the secondary electrons) and the tailing towards the high energy side [72-74].

However, one may consider another possibility where the slow electron is actively participating in the Auger process. The energy released during the filling of the inner shell vacancy is transferred to this slow electron and, as a result, the slow electron turns to be the "fast" Auger electron, leaving the singly charged ion. The energy spectrum of such "converted" electrons has to be very broad because of their initial continuous energy distribution (before the instant of filling the vacancy) and to some extent also, due to the lifetime of this vacancy in the neutral system, consisting of a singly-charged positive ion and a slowly outgoing electron. Observing this possible process by detecting these smeared energy electrons against the large background of the inelastically scattered or the ejected electrons (with a broad and continuous energy distribution) is not an easy task. An additional background contributions is usually not an experimental feature that can be detected accidentally. It is possible that due to this reason, the proposed process was not yet discussed in the literature. The probability of this process can be quite high, even for large distances between the atom and the slowly departing free electron, due to a strong overlap between wave functions of initial and final states of the outgoing (free) electron. This situation closely resembles the process of direct Auger relaxation of an excited atom in the vicinity of a metal surface (Auger de-excitation) [75,76]. Within this collisional process the energy released by the de-excitation of the atom into its ground state is transferred by an Auger mechanism to one of the conduction (nearly free) electrons of the metal. It was found [75,76] that for dipole transitions in the atom this process is very long range with a power law distance dependence ($R^{-4}$) of the transition rate (rather than the usual exponential one). Finally, one should also note that there are atoms, such as Si, Ga, Ge, In, Sn, Sb, Te, I, Xe, Hg, Pb, and Bi, for which inner shell contributions to $\sigma^+(E)$ with $E_{nl} \geq 100 \div 150$ eV calculated by Eq. (14) lead to overestimation, as compared with the experimental $\sigma^+(E)$ for the relevant (high) energy range of $E$ (see Fig. 1). One can assume that in these cases the Auger decay of the inner shell vacancies results in the production of multicharge ions. It seems that this is the case also for less deep inner shells (with $E_{nl} \approx 20 - 60$ eV) of atoms with outer $ns^2$ configurations such as Mg, Ca. Sr, Ba.

Some contribution to the integral single ionization cross-section $\sigma^+(E)$ can be given also by an autoionization process. The autoionizing states are created via excitation of more than one electron or via the formation of a vacancy in any inner shell followed by transition of the electron of this shell to one of the unoccupied states and an Auger relaxation resulting in a single charge ion production. For example, for the case discussed above of the Xe ionization, the question addressed is the creation of $N_{45}$ vacancy resulting in the configuration $4d^9 5s^2 5p^6 ns$ ($n \geq 6$). The lowest of these excited states is the $6p$ state whose excitation energy is 65 eV [45]. Indeed, the bend in the experimental $\sigma^+(E)$ dependence (see Fig. 1), which can be attributed to $N_{45}$ vacancies, appears at an energy value (E $\approx$ 60 eV) below the energy needed for the creation of $N_5$ vacancy (with direct ionization of the atom) which is 67.5 eV [45]. It is quite possible that in the range of 60 eV $\leq E \leq$ 65 eV, excitation of the Xe atom to the excited configuration $4d^9 5s^2 5p^6 ns$ ($n \geq 6$) takes place.

Kim et al. [4] associated the second maximum in the $\sigma^+(E)$ dependence (see Fig. 1d and Fig. 2) for a rubidium atom (ground state configuration $4s^2 p^6 5s$) with the autoionizing excitation of the $4p$ core electrons. Using the plane-wave Born approximation they have calculated the contribution of these autoionized states to the integral yield of Rb$^+$ ions. As can be seen from Fig. 2, in the range of E=40–100 eV the contributions of the autoionizing excitations $4p \rightarrow 4d$ or $5s$ or $5p$ to $\sigma^+(E)$ is 2.7 ÷1.4 times higher than direct core ionization ($4p^6$ and $4s^2$ shells). At the same time, the direct core ionization in this energy range is 19-65% of the 5s-shells direct ionization cross section.

Fig. 2 also presents the partial direct ionization cross section $\sigma^+_{5s}(E)$ and the total cross sections $\sigma^+(E) = \sigma^+_{5s}(E) + \sigma^+_{4p}(E) + \sigma^+_{4s}(E)$ as calculated by Eq. (14). It is interesting to note that $\sigma^+_{5s}(E)$ calculated by the BED theory [3] is in rather good agreement with the one calculated using Eq. (14), while the core cross section $\sigma^+_{4p}(E) + \sigma^+_{4s}(E)$ calculated by Eq. (14) is considerably higher than the one calculated by the BEB theory. Therefore, to achieve agreement with the experiment, the contribution of the core excitation has to be smaller than the one calculated within the plane wave Born approximation, as presented in Fig. 2. A very similar structure in the low energy region of the $\sigma^+(E)$ dependence exists also for the ionization of K and Cs (see Fig. 1).

## 6. Summary

Reliable models for the energy dependent single ionization cross-section $\sigma^+(E)$ of atoms by electron impact constitute a basic pre-requisite for any quantitative modeling of processes involving energetic electrons and neutral atoms or molecules. Unfortunately, it seems that currently available models are either very difficult to apply for most elements or lack predictive power. In this paper we have critically reviewed and analyzed some currently accepted approaches. We have then suggested a one-fitting-parameter expression for $\sigma^+(E)$ that was shown to be in good agreement with most of the measured cross-sections reported in the literature (from ionization threshold up to ionizing electron energy of 200 eV). The agreement obtained between our one-parameter expression as presented in this study and the experimental result, is at least as good as that obtained by other expressions using from several to tens of fitting parameters.

We propose that the single parameter as used in our $\sigma^+(E)$ expression is associated with the effective reduction in the number of equivalent electrons within a given atomic shell accessible for ionization by electron impact. For shells which are occupied by more than four electrons, this effective reduction was crudely modeled (classically) in terms of intrashell shadowing of part of the shell electrons by the other electrons in the same shell, while for $nl$-shells with $N_{nl} \leq 4$ our expression for the partial cross-section $\sigma_{nl}^+$ is parameter free. For elements with outer shell configuration $2s^22p^i$ ($i = 3,4,5,6$), namely N, O, F and Ne atoms the effective reduction of $N_{nl}$ is treated in terms of both intrashell and $sp$ intershell shadowing.

We have also discussed the contribution of inner shell ionization to the single ionization cross-section. For several atoms, taking into account single ionization of inner shells leads to better agreement between measured and calculated cross-sections. The better agreement is mainly manifested in the shape of the high-energy tail of the cross–section (100-200 eV). For other atoms, contributions due to deeper inner levels may lead to overestimation of the high-energy tail of the cross-section. We suggest that a variety of post-collision interactions (some of them already discussed in the literature) can explain the contribution of inner shell ionization to the single ionization cross-section. One possible mechanism involves the recapture of a slowly outgoing electron resulting in the formation of an excited singly charged core. The other process we describe is basically Auger de-excitation of an over-excited atom where a slow electron is actively participating in the Auger process.

**Acknowledgment**

We thank E. E. Nikitin for valuable discussions. This research was supported by the Israel Science Foundation (ISF) and by the James-Franck program.

# Appendix

There is some problem finding the binding energies for the inner $(n-1)s^2$ – electrons of alkali atoms with electron configurations $(n-1)s^2p^6ns$. Dirac – Fock calculations by Desclaux [44] of the binding energies for $(n-1)s^2$ – shells are larger by ~ 10.5 eV than the corresponding values given by Lotz [46] (see 5$^{th}$ column in Table A1). Here we shall estimate these binding energies using the following simple procedure. Let us consider a three stage process starting from a ground state alkali atom and leading to double charged ion in the excited configuration $(n-1)s\,p^6$ :

$$(n-1)s^2p^6ns \Rightarrow (n-1)s^2p^6 + e^- \Rightarrow (n-1)s^2p^5 + 2e^- \Rightarrow (n-1)sp^6 + 2e^- \ . \qquad (A1)$$

Energies for these transitions can be obtained from Moor's tables [45] and their sums for alkali atoms are presented at the 3$^{rd}$ column of Table A1. On the other hand, the excited state of doubly charged alkali ions with configuration $(n-1)sp^6$ can also be reached by an alternative two stage process:

$$(n-1)s^2p^6ns \Rightarrow (n-1)sp^6ns + e^- \Rightarrow (n-1)sp^6 + 2e^- \ . \qquad (A2)$$

The energy needed for the first transition in process (A2) is the binding energy of an electron in the inner $(n-1)s^2$ – shell, which is the one we are interested in. The second transition in (A2) is ejection of $ns$ – electron from single charged ion having vacancy in the $(n-1)s$ – shell. The energy needed for this transition is not known. However, one can expect that for an alkali element with atomic number $Z$ this energy is close to the one needed for ejection of $ns$ –electron from the singly charged ion of element with atomic number $Z+1$ and configuration $(n-1)s^2p^6ns$ . The basis for this assumption is that the effective charge (according to Slater's rules [77]) acting on the $ns$ – electron in alkali ion with atomic number $Z$ and configuration $(n-1)s\,p^6ns$  is very close to that in an alkaline earth ion with atomic number $Z+1$  and configuration $(n-1)s^2p^6ns$, namely 3.05 and 3.20 correspondingly. The IP values of ground state ions with atomic numbers $Z+1$ taken from Ref. [45] are presented in the 4$^{th}$ column of Table A1. Using these IP values instead of the energies needed for the second transition in the process (A2) together with the (A1) process energies given in the 3$^{rd}$ column in Table A1, one can calculate the energies needed for the first transition in the process (A2). These are the binding energies of the alkali atoms $(n-1)s^2$ electrons as listed in the 5$^{th}$ column. For the sake of comparison, the corresponding values from Desclaux [44] and Lotz [46] data are also presented at the same column.

Table A1

| 1 | 2 | 3 | 4 | 5 | | |
|---|---|---|---|---|---|---|
| $Atom_Z$ | Ground configuration | (A1) process energies | $Ion^+_{Z+1}$ (IP) | $E_{(n-1)s^2}$ | | |
| | | | | Present | Desclaux [44] | Lotz [46] |
| $Na_{11}$ | $2s^2p^63s$ | 85.2 | $Mg^+_{12}$ (15.0) | 70.2 | 76.3 | 66 |
| $K_{19}$ | $3s^2p^64s$ | 52.3 | $Ca^+_{20}$ (11.9) | 40.5 | 47.9 | 37 |
| $Rb_{37}$ | $4s^2p^65s$ | 48.1 | $Sr^+_{38}$ (11.0) | 37.1 | 42.6 | 32 |
| $Cs_{55}$ | $5s^2p^66s$ | 44.8 | $Ba^+_{56}$ (10.0) | 34.8 | 35.6 | 25 |

Caption for Table A1: Electron binding energies related with the (A1) and (A2) processes for alkali atoms. The energy for the generation of an excited doubly charged ion ((A1) process) is given in the 3$^{rd}$ column. The IP energies for the ground state alkaline earth ions with atomic numbers $Z+1$ (see text) are given in the 4$^{th}$ column. The presently evaluated binding energies of inner $(n-1)s^2$ electrons of alkali atoms $E_{(n-1)s^2}$ are given in the 5$^{th}$ column together with Desclaux and Lotz values. All energies are given in eV.

**References**

1.  M. R. H. Rudge, Rev. Modern Phys. 40, 564 (1968).
2.  T. D. Märk, G.H. Dunn (Eds.), Electron Impact Ionization, Springer Verlag, Viena, 1985
3.  Y.-K. Kim and M. E. Rudd, Phys. Rev. 50, 3954 (1994).
4.  W. Hwang, Y.-K. Kim and M. E. Rudd, J. Chem. Phys. 104, 2956 (1996).
5.  Y.-K. Kim, W. Hwang, N. M. Weinberger, M. A. Ali, and M. E. Rudd, J. Chem. Phys. 106,1026 (1997).
6.  Y.-K. Kim, J. Migdalek, W. Siegel and J. Bieroń, Phys. Rev. A57, 246 (1998).
7.  Y.-K. Kim and P.M. Stone, Phys. Rev. A64, 052707(2001).
8.  D. Margreiter, H. Deutsch and T. D. Märk, Int. J. Mass Spectrom. Ion Processes139, 127 (1994).
9.  H. Deutsch, K. Becker and T. D. Märk, Contr. Plasma Phys. 35, 421 (1995).
10. H. Deutsch, K. Becker and T. D. Märk, Int. J. Mass Spectr. 177, 47 (1998).
11. H. Deutsch, K. Becker and T. D. Märk, Int. J. Mass Spectr. 185/185/187, 319 (1999).
12. H. Deutsch, K. Becker, S. Matt, and T. D. Märk, Int. J. Mass Spectr. 197, 37 (2000).
13. J.-M. Rost, J. Phys. B: At. Mol. Opt. Phys. 28, 3003 (1995).
14. V. Saksena, M. S. Kushwaha and S. P. Khare, Physica B23, 201 (1997).
15. S. P. Khare, M. K. Sharma and S. Tomar, J. Phys. B: At. Mol. Opt. Phys. 32, 3147 (1999).
16. K. N. Joshipura and B. K. Antony, Phys. Lett. A289, 323 (2001).
17. K. N. Joshipura and C.G. Limbachiya, , Int. J. Mass Spectr. 216, 239 (2002).
18. R. S. Freund, R. C. Wetzel, R. J. Shul and T. R. Hayes, Phys. Rev. A41, 3575 (1990).
19. P. McCallion, M.B. Shah and H.B. Gilbody, J. Phys. B: At. Mol. Opt. Phys. 25, 1051 (1992).
20. P. McCallion, M.B. Shah and H.B. Gilbody, J. Phys. B: At. Mol. Opt. Phys. 25, 1061 (1992).
21. B. Min, Y. Yoshinari, T. Watabe, Y. Tanaka, C. Takayanagi, T. Takayanagi, K. Wakiya and M. Suzuki, J. Phys. Soc. Jap. 62, 1183 (1993).
22. A. R. Johnson and P. D. Burrow, Phys. Rev. A51, R1735 (1995) .
23. K. Fujii and S. K. Srivastava, J. Phys. B: At. Mol. Phys. 28, L559 (1995).
24. R.S. Schappe, T. Walker, L.W. Anderson and C.C. Lin, Phys. Rev. Lett. 76, 4328 (1996).
25. C. J. Patton, K. O. Lozhkin, M. B. Shan, J. Geddes and H.B. Gilbody, J. Phys. B: At. Mol. Opt. Phys. 29, 1409 (1996).
26. D. P. Almeida, M. A. Scopel, R. R. Silva and A. C. Fontes, Chem. Phys. Lett. 341, 490 (2001).
27. R. Rejoub, B. J. Lindsay and R. F. Stebbings Phys. Rev. A65, 042713 (2002).

28. J. J. Thomson, Phys. Mag. 23, 449 (1912).

29. M. Gryzinski, Phys.Rev. A138, 305 (1965); *ibid* A138, 322 (1965); *ibid* A138, 336 (1965).

30. R. C. Stabler, Phys. Rev. 133, 1268 (1964).

31. E. Gerjuoy, Phys. Rev. 148, 54 (1966).

32. A. Burguess and I.C. Percival, Adv. Atom. Mol. Phys. 4, 109 (1968).

33. A. Burguess, Proc. Symp. Atomic Collision Processes in Plasmas: Culham, 1964, p. 63.

34. V. I. Ochkur, Proc. Leningrad State Univers. "The Problems of Atomic Collision Theory" No. 1, p. 42 (1975) (in Russian).

35. L. Vriens, Phys. Rev. 141, 88 (1966).

36. L. Vriens, Proc. Phys. Soc. 89, 13 (1966).

37. L. Vriens, in Case Studies in Atomic Physics, edited by E. W. McDaniel and M. R. C. McDowell (North-Holland, Amsterdam, 1969), Vol. 1, p. 335.

38. N. F. Mott, Proc. Roy. Soc. London, Ser. A126, 259 (1930).

39. L. D. Landaw and E. M. Lifshitz, Quantum Mechanics, Non relativistic Theory, 2nd ed. (Addison-Wesley, Reading, MA, 1965), p. 575.

40. H. Bethe, Ann. Physik 5, 325 (1930).

41. J. B. Mann, J. Chem. Phys. 46, 1646 (1967).

42. Y.-K. Kim and M.E. Rudd, J. Phys. B: At. Mol. Opt. Phys. 33, 1981 (2000).

43. H. Deutsch and T. D. Märk, Int. J. Mass Spectrom. Ion Processes 79, R1 (1987).

44. J. P. Desclaux, Atomic Data and Nuclear Data Tables 12, 311 (1973).

45. C. E. Moore, Atomic Energy Levels, CNBS 467, 1952.

46. W. Lotz, J. Opt. Soc. Am. 60, 206 (1970).

47. W. L. Fite and R. T. Brackmann, Phys. Rev. 112, 1141 (1958).

48. E. W. Rothe, L. L. Marino, R. H. Neynaber and S. M. Trujillo, Phys. Rev. *125*, 582 (1962).

49. M. B. Shah, D. S. Elliot and H. B. Gilbody, J. Phys. B: At. Mol. Phys. 20, 3501 (1987).

50. E. Brook, M.F.A. Harrison and A.C.H. Smith, J. Phys. B: At. Mol. Phys. 11, 3115 (1978).

51. K. Stephan, H. Helm and T. D. Märk, J. Chem. Phys. 73, 3763 (1980).

52. R. C. Wetzel, F. A. Baiocchi, T. R. Hayes and R. S. Freund, Phys. Rev. A35, 559 (1987).

53. R. H. McFarland and J. D. Kinney, Phys. Rev. A137, 1058 (1965).

54. I. P. Zapesochnyi and I. S. Aleksakhin, Sov. Phys. JETP 28, 41 (1969).

55. R. Jalin, R. Hagemann and R. Botter, J. Chem. Phys. 59, 952 (1973).

56. T. R. Hayes, R. C. Wetzel and R. S. Freund, Phys. Rev. A35, 578 (1987).

57. J. T. Tate and P. T. Smith, Phys. Rev. 46, 773 (1934).

58. L. A. Vainshtein, V. I. Ochkur, V. I. Rakhovskii and A. M. Stepanov, Zh. Eksp. Teor.Fiz. 61, 511 (1971) (Sov. Phys. JETP 34, 271 (1972)).

59. D. G. Golovach, A. N. Drozdov, V. I. Rakhovskii and V. M. Shustryakov, Izmer. Tekh. 6, 51 (1987) [Meas. Tech. (USSR) 30, 587 (1987)].

60. S. Okudaira, J. Phys. Soc. Jap. 29, 409 (1970).

61. L. A. Vainstein, D. G. Golovach, V. I. Ochkur, V. I. Rakhovskii, N. M. Rumyantsev, and V. M. Shustryakov, Zh. Eksp. Teor. Fiz. 93, 65 (1987), Sov. Phys. JETP 66, 36 (1987).

62. K. J. Nygaard and Y. B. Hahn, J. Chem. Phys. 58, 349 (1973).

63. K. Stephan and T. D. Märk, J. Chem. Phys. 81, 3116 (1984).

64. L. L. Shimon, P. N. Volovich and M. M. Chiriban, Zh. Tekh, Fiz. 59, 64 (1989) (Sov. Phys. Tech. Phys. 34, 1264 (1989)).

65. W. Bleakney, Phys. Rev. 35, 139 (1930).

66. S. L. Pavlov and G .I. Stotskii, Zh. Eksp. Teor. Fiz. 58, 108 (1970), Sov. Phys. JETP 31, 61 (1970).

67. J. C. Halley, H. H. Lo and W. L. Fite, Phys. Rev. A23, 1708 (1981).

68. O. Keski-Rakhonen and M.O. Krause, Atomic Data and Nuclear Data Tables 14, 139 (1974).

69. T. Takayanagi, A. Nakashio, C. Hirota, H. Suzuki, and K. Wakiya, Proc. XI ICPEAC, Abstract of papers, eds. K. Takayanagi and N. Oda (The Society for Atomic Collision Research, Japan, 1979) p. 260.

70. Y. Yagishita, Phys. Lett. 87A, 30 (1981).

71. H. Suzuki, T. Hirayama, T. Takayanagi, Proc. XVI ICPEAC, New York 1989 (AIP Proc. 205, 1990) p. 82.

72. S. Ohtani, H. Nishimura, H. Suzuki, and K. Wakiya, Phys. Rev. Lett. 36, 863 (1976).

73. A. Niehaus, J. Phys. B: Atom. Mol.Phys. 10, 1845 (1977).

74. D. Čubrić, A. A. Wills, E. Sokell, J. Comer and M. A. MacDonald, J. Phys. B: At. Mol. Opt. Phys. 26, 4425 (1993).

75. V. Pazdzersky and B. Tsipinyuk, Abstracts of All-Union Seminar "Ion Beam Diagnostics of Surfaces", Uzhgorod, 1985, p. 144 (in Russian).

76. V. Pazdzersky and B. Tsipinyuk, The Proc. Of VII All-Union Conf. On Interaction of Atomic Particles with Solid, Moskva, 1987, v. 3, p. 10. (in Russian).

77. J. C. Slater, Quantum Theory of Atomic Structure, V.1, 2, London-New York-Toronto, McGraw-Hill Book Company Inc., 1960.

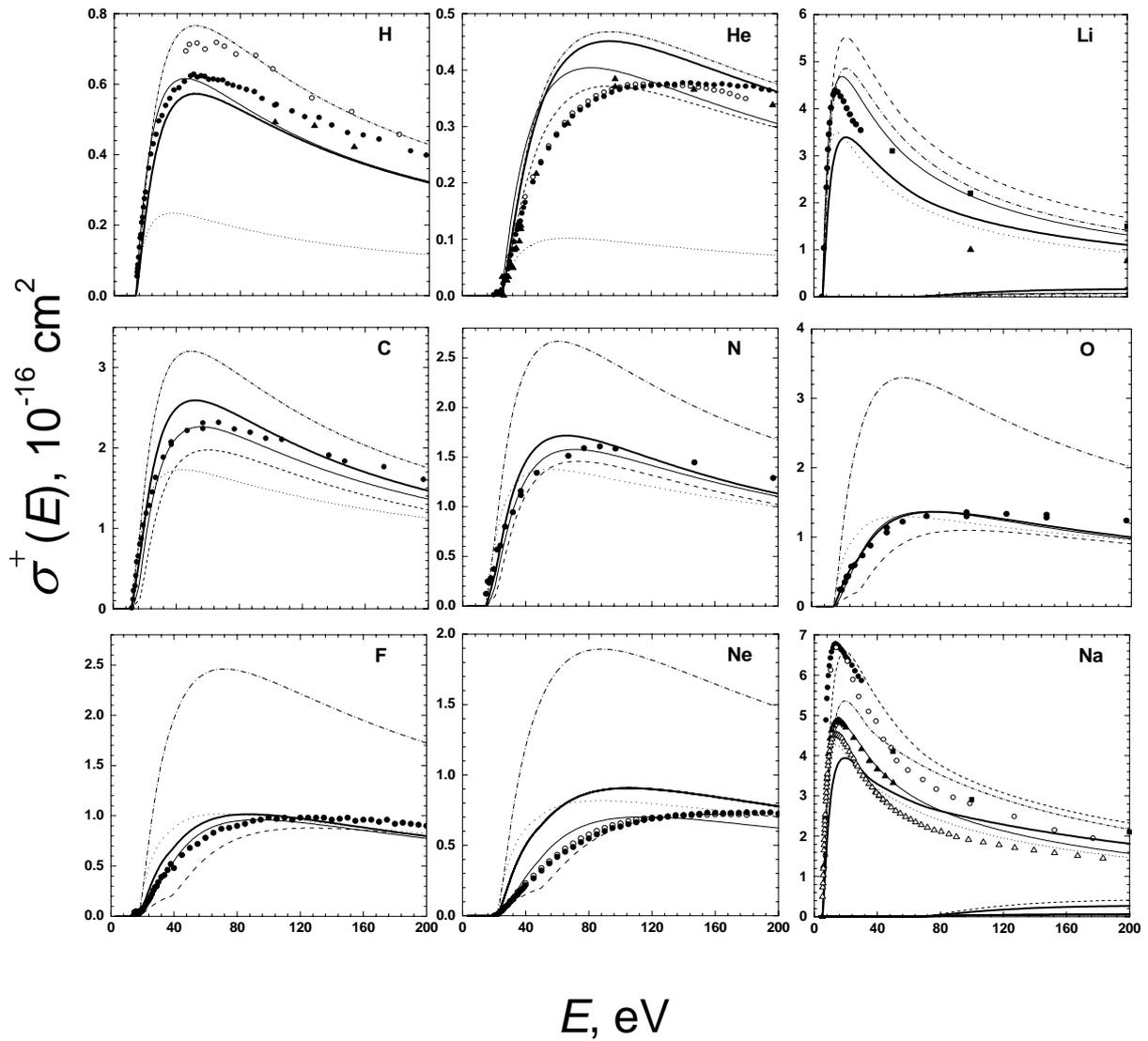

**Fig 1(a)** Electron impact single ionization cross-section $\sigma^+$ vs electron energy E. The $\sigma^+(E)$ dependencies as calculated by formula (14) are given by the thick solid line; by Gryzinski's formulae (4), (5) – by the dash-dot line; by Mann's formula (9) – by the dotted line; by DM formula (10) – by the dashed line and by DM formula (13) – by the thin solid line. For Li and Na also the partial cross-sections $\sigma^+_{1s}(E)$ and $\sigma^+_{2s}(E)$ (correspondingly) are also presented. Note that for hydrogen the calculations by formula (14) and by the DM formula (10) give the same results. This is due to the fact that using $g_s=3$ in formula (10) for ground state hydrogen leads to $g_s r^2_{max} = \langle r^2 \rangle$. The experimental data for H atom are from Fite and Brackman [47] (○), from Rothe et al. [48] (▲) and from Shah et al. [49] (●); for He - from Brook et al. [50] (▲), from Stephan et al. [51] (○) and from Wetzel et al. [52] (●); for Li - from McFarland and Kinney [53] (○) (gross $\sigma^+$), from Zapesochnyi and Aleksakhin [54] (●) (gross $\sigma^+$) and from Jalin et al. [55] (▲); for C, N and O - from Brook et al. [50] (●); for F - from Hayes et al. [56] (●); for Ne - from Stephan et al. [51] (○) and from Wetzel et al. [52] (●); for Na - from McFarland and Kinney [53] (■) (gross $\sigma^+$), from Zapesochnyi and Aleksakhin [54] (●) (gross $\sigma^+$), from Johnston and Burrow [22] (▲), from Fujii and Srivastava [23] (Δ) and from Tate and Smith [57] (○) ( relative $\sigma^+(E)$ dependence, normalized to the data given in Ref. [54] at E = 14 eV).

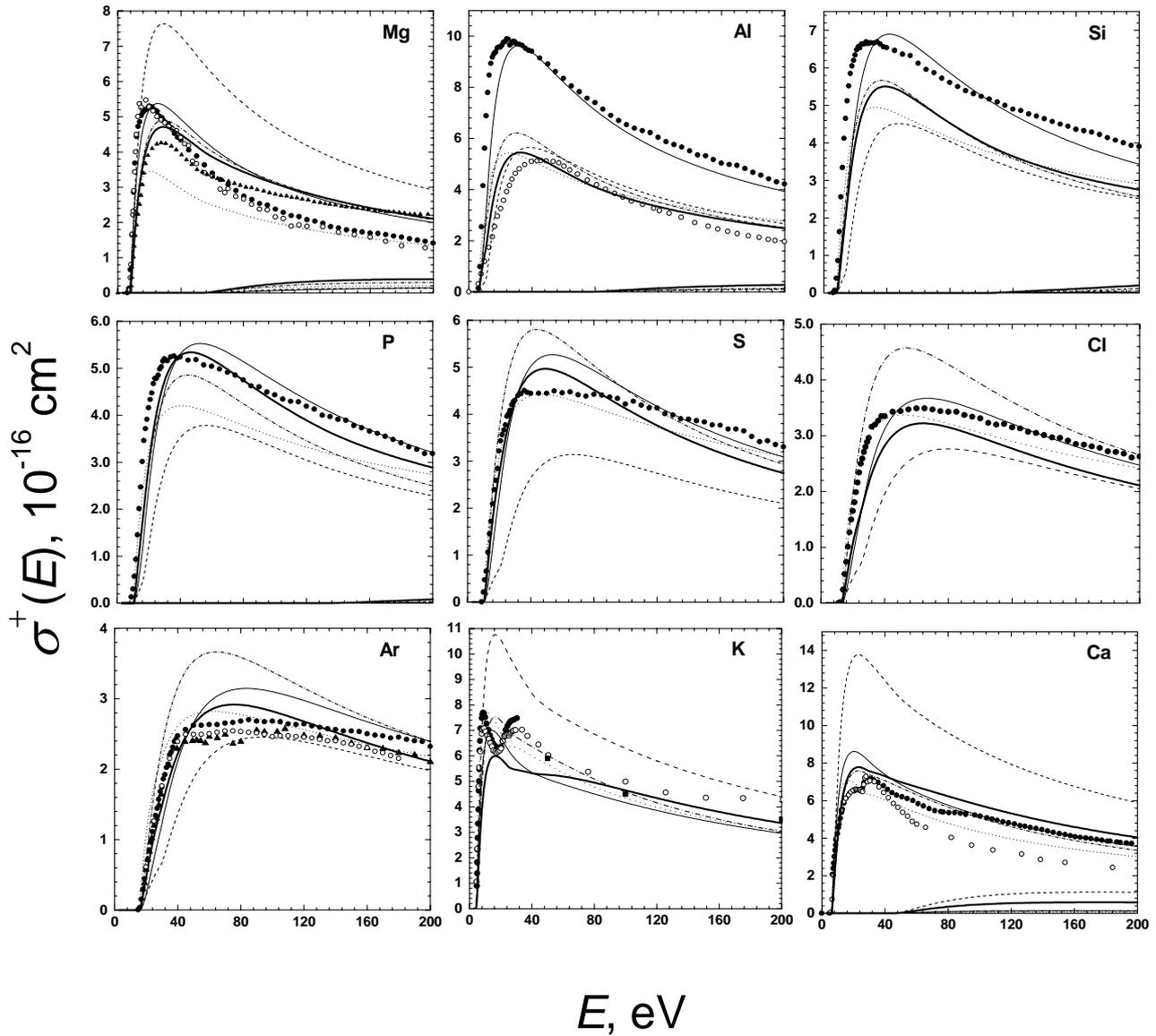

**Fig. 1 (b)** Electron impact single ionization cross-section $\sigma^+$ vs electron energy E. The $\sigma^+(E)$ dependences as calculated by formula (14) are given by the thick solid line, by Gryzinski formulae (4), (5) – by the dash-dot line, by Mann formula (9) – by the dotted line, by DM formula (10)- by the dashed line and by DM formula (13) – by the thin solid line. The partial cross-sections $\sigma_{2p}^+$ for Mg, Al, Si and P, and $\sigma_{3s}^+$ for Ca are also presented. The experimental data for Mg atom are from Freund et al. [18] (●), from P. McCallion et al. [19] (○) and from Vainstein et al. [58] (gross $\sigma$) (▲); for Al - from Freund et al. [18] (●) and from Golovach et al. [59] (○) (gross $\sigma^+$); for Si, P and S - from Freund et al. [18]; for Cl - from Hayes et al. [56]; for Ar - from Stephan et al. [51] (○), from Wetzel et al. [52] (●) and from P.McCallion et al., [20] (▲); for K - from McFarland and Kinney [53] (■) (gross $\sigma^+$), from Zapesochnyi and Aleksakhin [54] (●) (gross $\sigma^+$) and from Tate and Smith [57] (○) (relative $\sigma^+(E)$ dependence, normalized to the data from Ref.[54] at $E=20$ eV ); for Ca - from Vainshtein et al. [58] (●) (gross $\sigma^+$) and from Okudaira [60] (○) (relative $\sigma^+(E)$ dependence, normalized to the data from Ref. [58] at E=20 eV).

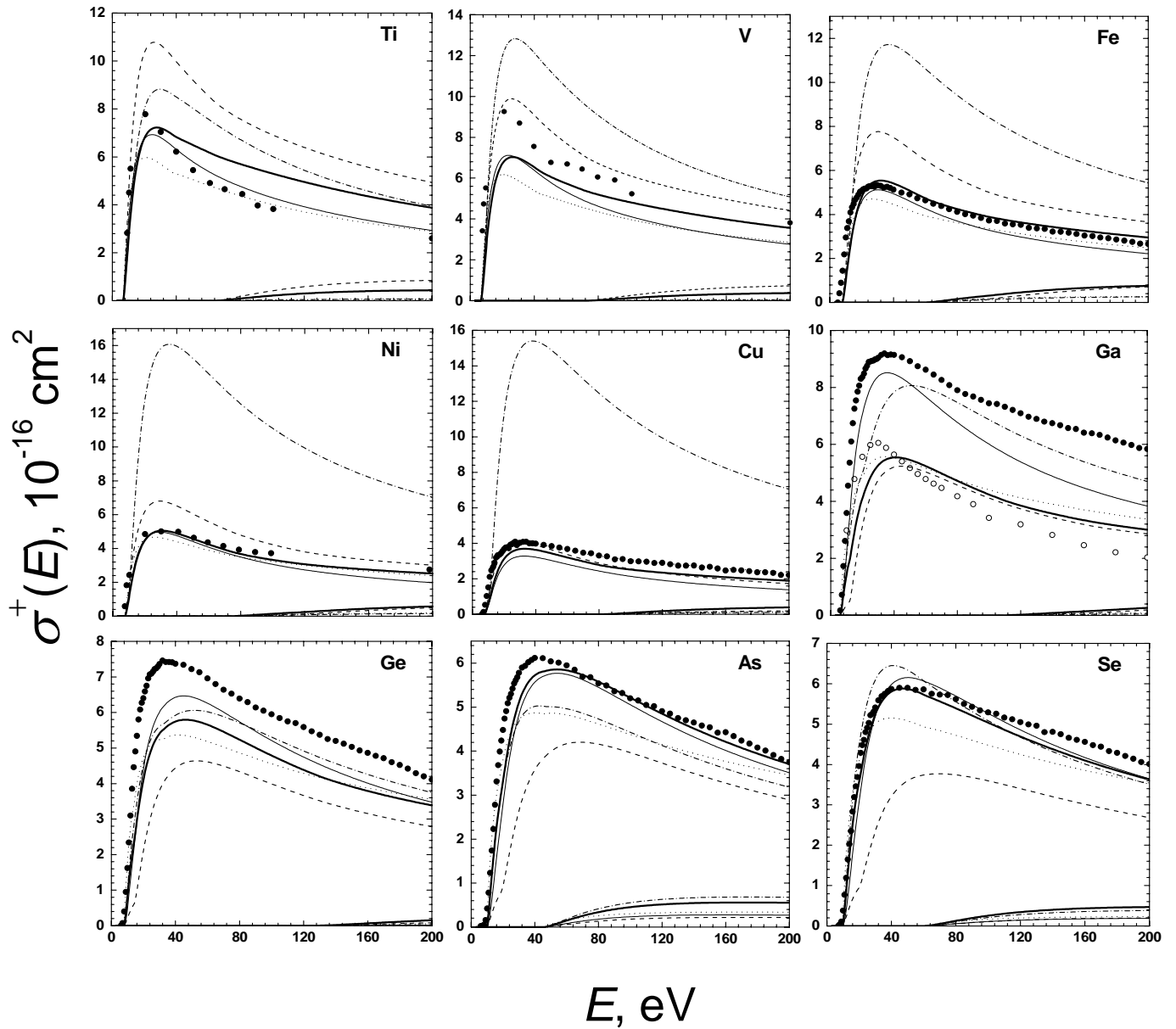

**Fig. 1 (c)**. Electron impact single ionization cross-section $\sigma^+$ vs electron energy E. The $\sigma^+(E)$ dependencies as calculated by formula (14) are given by the thick solid line; by Gryzinski's formulae (4), (5) – by the dash-dot line; by Mann formula (9) – by the dotted line, by DM formula (10) – by the dashed line and by DM formula (13) – by the thin solid line. The partial cross-sections $\sigma^+_{3s}$ for Ti and V, ($\sigma^+_{3p} + \sigma^+_{3s}$) for Fe, Ni, Cu and Ga, $\sigma^+_{3p}$ for Ge and $\sigma^+_{3d}$ for As and Se also are presented. The experimental data for Ti, V and Ni are from Koparnski as presented in Ref. [8]; for Fe, Cu, Ge, Ga, As and Se - from Freund et al. [18](●) and for Ga - from Vainshtein et al. [61] (○).

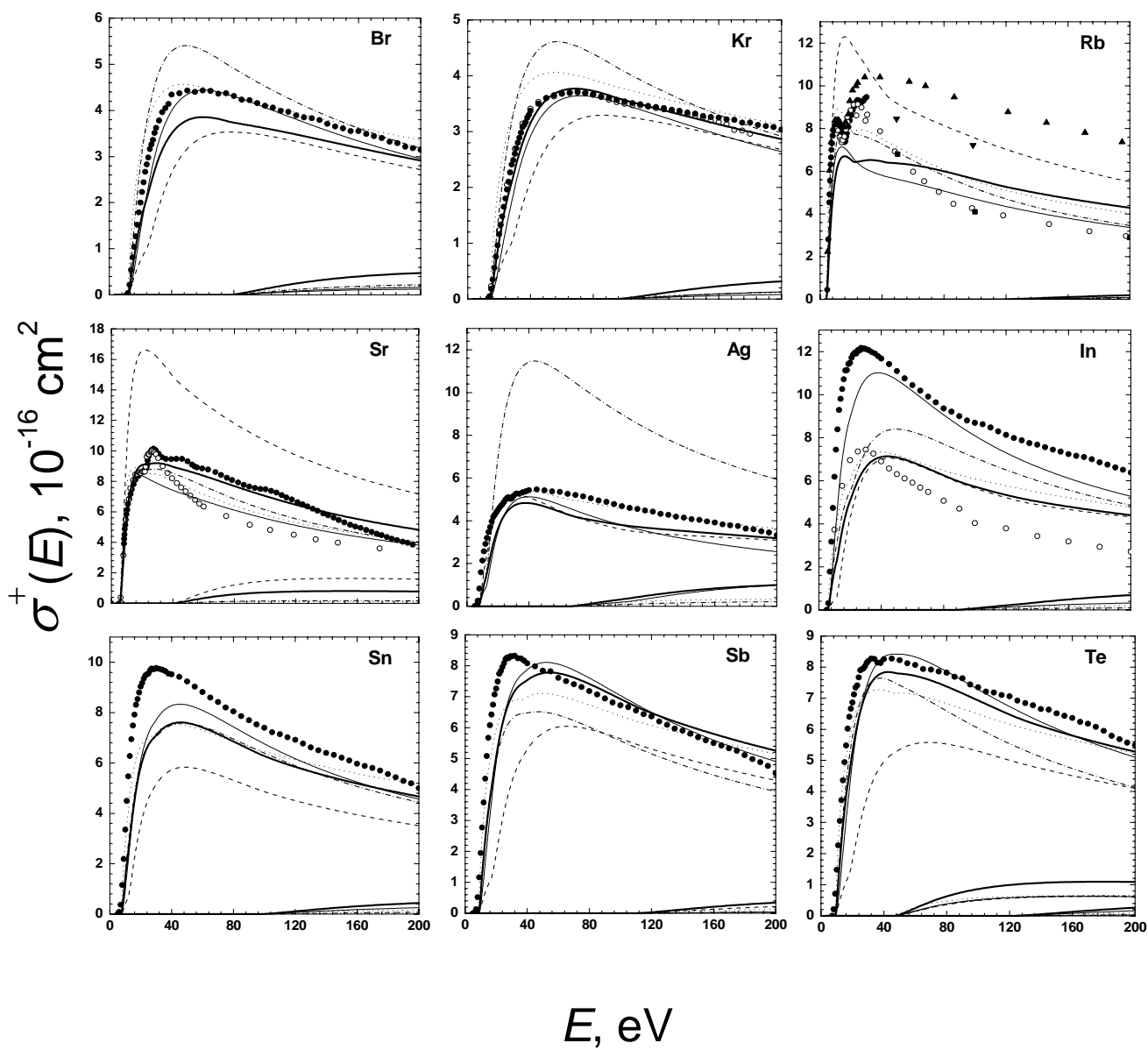

**Fig. 1 (d)** Electron impact single-ionization cross-section vs. electron energy E. The $\sigma^+(E)$ dependencies as calculated by formula (14) are given by the thick solid line; by Gryzinski's formulae (4), (5) – by the dash-dot line; by Mann's formula (9) – by the dotted line; by DM formula (10) – by the dashed line and by DM formula (13) – by the thin solid line. The partial cross-sections $\sigma^+_{3d}$ for Br, Kr and Rb, $\sigma^+_{4s}$ for Sr, ($\sigma^+_{4p}+\sigma^+_{4s}$) for Ag and In, $\sigma^+_{4p}$ for Sn, Sb and $\sigma^+_{4d}$ and $\sigma^+_{4p}$ for Te are also presented. The experimental data for Br atom are from Hayes et al. [56]; for Kr - from Stephan et al. [51] (○) and from Wetzel et al. [52] (●); for Rb - from Zapesochnyi and Aleksakhin [54] (●) (gross $\sigma^+$), from Nygaard and Hahn [62] (▲) (gross $\sigma^+$), from Tate and Smith [57] (○) (relative $\sigma^+(E)$ dependence, normalized to the data from Ref. [38] at E=10eV), from McFarland and Kinney [53] (■) (gross $\sigma^+$) and from Schappe et al. [24] (▼); for Sr - from Vainstein et al. [58] (●) (gross $\sigma^+$) and from Okudaira [60] (○) (relative $\sigma^+(E)$ dependence, normalized to the data from Ref. [58] at E=25 eV); for Ag and In - from Freund et al. [18] (●) and Vainshtain et al. [61] (○) (gross $\sigma^+$); for Sn, Sb and Te - from Freund et al. [18].

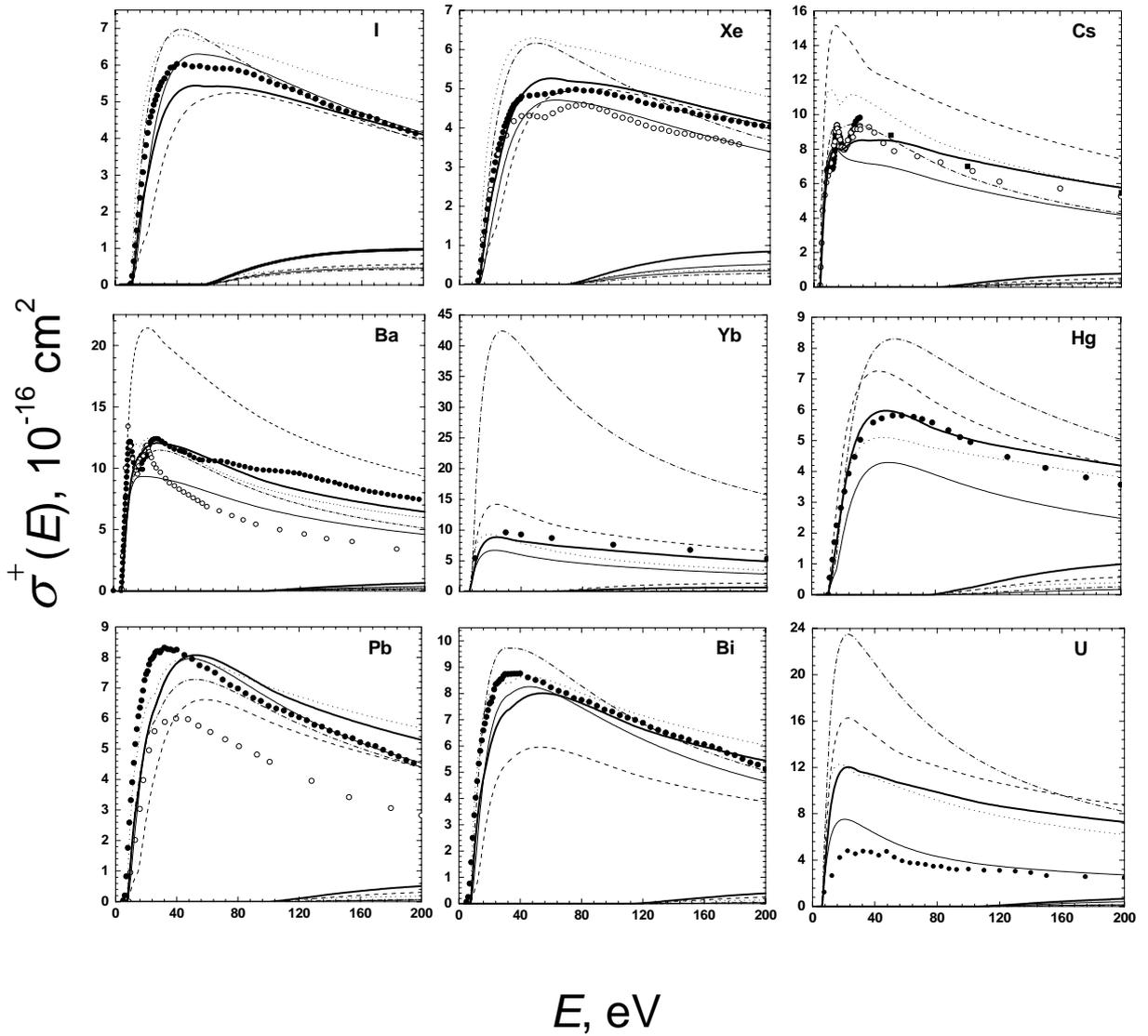

**Fig. 1 (e)** Electron impact single ionization cross-section $\sigma^+$ vs electron energy E. The $\sigma^+(E)$ dependencies as calculated by formula (14) are given by the thick solid line; by Gryzinski formulae (4), (5) – by the dash-dot line; by Mann formula (9) – by the dotted line; by DM formula (10) – by the dashed line and by DM formula (13) – by the thin solid line. The partial cross-sections $\sigma^+_{4d}$ for I, Xe, Cs and Ba, $\sigma^+_{5s}$ for Yb, ($\sigma^+_{5p} + \sigma^+_{4f}$) for Hg, $\sigma^+_{5p}$ for Pb and Bi and $\sigma^+_{5d}$ for U are also presented. The experimental data for I atom are from Hayes et al. [56]; for Xe - from Wetzel et al. [52] (●) and from Stephan and Mark [63] (○); for Cs - from Zapesochnyi and Aleksakhin [38] (●) (gross $\sigma^+$), from McFarland and Kinney [37] (■) (gross $\sigma^+$) and Tate and Smith [57] (○) (relative $\sigma^+(E)$ dependence, normalized to data from Ref.[54] at E = 20eV); for Ba - from Vainshtein et al. [58] (●) and from Okudaira [60] (○) (relative $\sigma^+(E)$ dependence, normalized to data of Ref.[58] at E = 20eV); for Yb - from Shimon et al. [64], for Hg - from Bleakney [65]; for Pb - from Freund et al. [18] (○) and from Pavlov and Stotskii [66] (●), for Bi - from Frend et al. [18] and for U - from Halley et al. [67].

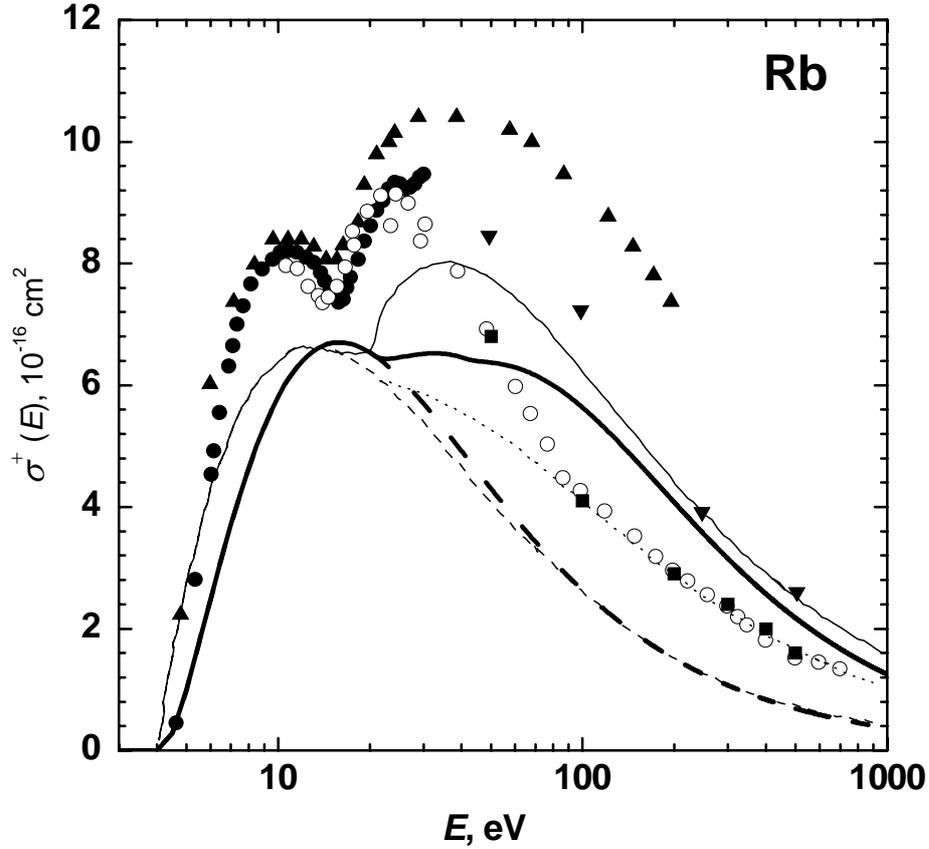

**Fig. 2.** Electron impact single ionization cross-section $\sigma^+(E)$ calculated for the Rb atom. The thick solid and dashed lines represent the total single ionization cross-section $\sigma^+(E)$ and partial $\sigma^+_{5s}(E)$ cross-section as calculated by formula (14) and by the 5s-term in this formula correspondingly. The total single ionization cross-section $\sigma^+(E)$ calculated by Kim et al. [6] is given by the thin solid line, while the partial $\sigma^+_{5s}(E)$ (BED theory [3,6]) is given by the thin dashed line. The difference between the thin dashed line and the dotted line represents the contribution of direct ionization of core electrons and the difference between the dotted line and the thin solid line represents the contribution of autoionization of Rb atoms with excited 4p-core electrons [6]. In order to properly compare between $\sigma^+(E)$ calculations by the different approaches, all $E_{nl}$ values, used in these calculations, were taken from Ref. (6). The experimental data are from Zapesochnyi and Aleksakhin [54] (●) (gross $\sigma^+$), from Nygaard and Hahn [62] (▲) (gross $\sigma^+$); from Tate and Smith [57] (○) (relative $\sigma^+(E)$ dependence, normalized to data from Ref. [54] at E=10eV), from McFarland and Kinney [53] (■) (gross $\sigma^+$) and from Schappe et al. [24] (▼).